\documentclass[a4paper,fleqn]{cas-sc}
\usepackage[authoryear,longnamesfirst]{natbib}
\usepackage{comment}
\usepackage{float}

\def\tsc#1{\csdef{#1}{\textsc{\lowercase{#1}}\xspace}}
\tsc{WGM}
\tsc{QE}

\begin{document}
\let\WriteBookmarks\relax
\def\floatpagepagefraction{1}
\def\textpagefraction{.001}

\shorttitle{Renewable Forecasting using Foundation Models}    

\shortauthors{}  

\title [mode = title]{Empirical Assessment of Time-Series Foundation Models For Power System Forecasting Applications}  



%

\author[1]{Muhy Eddin Za'ter}

\cormark[1]


\ead{muhy.zater@colorado.edu}



\affiliation[1]{organization={University of Colorado Boulder},
            addressline={address}, 
            city={Boulder},
            postcode={80303}, 
            state={Colorado},
            country={USA}}

\author[1,2]{Bri-Mathias Hodge}


\ead{brimathias.hodge@colorado.edu}



\affiliation[2]{organization={National Renewable Energy Laboratory},
            addressline={address}, 
            city={Golden},
            postcode={80302}, 
            state={Colorado},
            country={USA}}

\cortext[1]{Corresponding author}



\begin{abstract}
Accurate forecasting of electric load and renewable generation is essential for reliable and cost‐effective power system  operations. Recent advances in transformer‐based and foundation machine learning models—driven by large‐scale pretraining, increased available data and computation, in addition to architectural innovations— have shown promise in time‐series forecasting across multiple domains. However, their application to power system forecasting tasks remains  largely underexplored. This work presents a comprehensive, empirical benchmark of state‐of‐the‐art time‐series foundation models, transformer architectures, and deep learning baselines for solar, wind, and load forecasting using the high‐resolution ARPA‐E PERFORM dataset for the Electric Reliability Council of Texas (ERCOT) grid. Eight core capabilities are assessed, including zero‐shot performance, fine‐tuning efficiency, multivariate input and output handling, horizon sensitivity, generalization to unseen sites, probabilistic forecasting, and context window effects. Models evaluated include TimesFM, Chronos‐Bolt, Moirai‐L, MOMENT, Tiny Time Mixer, Temporal Fusion Transformer, PatchTST, TimeXer, LSTM, and CNN. The manuscript aims to provide clear guidance on when foundation models can provide enhanced renewable and load forecasting capabilities and when other approaches remain the more practical choice for power system operations.
\end{abstract}



\maketitle

\section{Introduction}

Power system planning and operations involve a broad spectrum of tasks and decisions that are typically determined ahead of the real-time action, and therefore, often rely on forecasts of various horizons and resolution \cite{xifan1994modern}. These decisions span from short-term actions like unit commitment and economic dispatch to longer-term strategies such as maintenance scheduling and multi-year capacity and infrastructure expansion. This is due in part to the physical limitations of the system; for instance many thermal generators have considerable start-up and ramping constraints, while building new generation and transmission infrastructure can take years \cite{soder2011efficient}. As a result, the quality of decisions made across timescales, from sub-minute frequency regulation to long-term resource adequacy planning depends critically on the accuracy of the underlying forecasts.

The importance of accurate forecasting has only increased with the widespread integration of variable renewable energy (VRE) sources, chiefly wind and solar power generation \cite{sweeney2020future}. These weather-dependent technologies introduce significant uncertainty into the system, affecting a wide spectrum of operational decisions. Studies have shown that even modest improvements in solar and wind forecasting can reduce system operating costs \cite{martinez2016value, wang2016quantifying}, while better demand forecasts can lower the need for balancing reserves, and reduce reliance on costly peaker plants, and enhance transient stability margins \cite{kuster2017electrical}. Thus, forecasting accuracy plays a central role in minimizing overall system costs and improving reliability, making it a key area of ongoing power systems research \cite{davis2024impact}.

In the past decade machine learning (ML) and deep learning (DL) methods, such as gradient-boosted trees, convolutional neural networks, and recurrent neural networks, have outperformed classical statistical models (i.e ARIMA models) in different forecasting tasks such as solar and wind generation, and electrical load \cite{shamshirband2019survey, miller2024survey}. Despite this progress, large-scale studies still identify four main challenges that limits the widespread adoption of these machine-learning based algorithms. First, being data hungry, these models tend to generalize poorly to new sites or regions with limited data \cite{paletta2021benchmarking}. Second, they often struggle with multivariate inputs and dependencies \cite{yin2023assessing}. Third, many architectures face scalability issues when dealing with long historical input sequences or multi-horizon prediction targets \cite{shringi2025review}. Finally, effective deployment to production and real-world sites often requires significant task specific tuning \cite{depoortere2025solnet}.

These limitations have sparked growing interest in transformer-based architectures and foundation models that have fundamentally revolutionized fields like natural language processing and computer vision \cite{islam2024comprehensive, myers2024foundation} and are making their way into time-series applications \cite{liang2024foundation}. Several developments make such models particularly promising for power system forecasting tasks: (i) the emerging availability of massive, multi-domain time-series corpora at the petabyte scale, which helps offset the challenge of limited site-specific data during pre-training \cite{tavenard2020tslearn} (ii) the development of advanced machine learning algorithms capable of effectively learning from these massive datasets, (iii) the increasing accessibility of large-scale training compute infrastructure via cloud-based Graphical and Tensor Processing Units (GPUs/TPUs) clusters \cite{sevilla2022compute}, and (iv) the demonstrated success of transformer-based sequence models in achieving state-of-the-art performance in zero-shot (without any training) and few-shot learning (training with limited samples) across diverse domains \cite{brown2020language}. Therefore, researchers have been developing and investigating the potential of these models (either general Large Language Models or time series transformer and foundation models) for time series tasks such as anomaly detection, data imputation, or forecasting \cite{liang2024foundation, ye2024survey}. 
In recent years more than 50 time series foundation models have been proposed for all forecasting domains, with different architectures, domains and applications \cite{liang2024foundation, kottapalli2025foundation}.

However, the evidence regarding the applicability of transformers and time series foundation models for forecasting is still mixed and the effectiveness of self attention and transformers are still in question \cite{zeng2023transformers, kim2024self}. Rigorous ablation studies show that simple linear baselines can outperform many transformer variants once strong seasonality cues are present in the dataset, calling into question the actual incremental value of the different attention mechanisms for time series \cite{xu2024specialized}. On the other hand, other experiments on manufacturing and macro‑economic datasets report significant gains for transformer-based models over even well‑tuned statistical and other machine learning models \cite{yeh2023toward, niu2024attention}. Therefore, there is still a lack of consensus in the community about the effectiveness of transformers and self-attention for time-series forecasting.
Given the wide spectrum of application domains for time series forecasting and the large number of variations in transformer-based and time-series foundation models, the variation in performance and effectiveness across different models and domains is a natural consequence. 

There have been surprisingly few attempts at exploring the potential of foundation and transformer models within the power and energy forecasting communities. 
For instance, \cite{meftahul2025foundation} provided a comprehensive review compiling the performance gains of various foundation models in the clean energy sector, drawing on the results reported in the models' original publications for clean energy forecasting applications.
In another notable example \cite{meyer2024benchmarking} benchmarked the performance of time series foundation models for household electricity load forecasting. Whereas the work in \cite{saravanan2024analyzing} focused on analysis of bulk load forecasting performance using TimeGPT, which is a proprietary model.
However, the power system community lacks a comprehensive and empirical study of the potential for foundation models and self-attention mechanisms for their most significant forecasting tasks \cite{zhao2026review}.

This paper presents a comprehensive, head-to-head evaluation of state-of-the-art time-series transformer and foundation models on power grid forecasting applications. It leverages the publicly available ARPA-E PERFORM dataset \cite{bryce2023solar, bryce2024solar}, which offers 5-minute resolution actuals and probabilistic forecasts for load, wind, and solar across the major United States Independent System Operator (ISO) the Electric Reliability Council of Texas (ERCOT). We benchmark these models across eight core capabilities where such models and architectures are expected to excel. For comparison, we also include conventional deep learning baselines, specifically Long Short-Term Memory (LSTM) networks and Convolutional Neural Networks (CNNs).

The evaluation focuses on the following capabilities:

\begin{itemize}
\item \textbf{Zero-Shot Forecasting:} Assessing performance using off-the-shelf, pre-trained models without any fine-tuning or task-specific adaptation.
\item \textbf{Fine-Tuning Capability:} Evaluating data efficiency when limited site-specific samples are available.
\item \textbf{Multivariate Input Handling:} Evaluating the model's ability to process and leverage multiple input features—such as weather variables, historical load, wind, and solar data—for accurate forecasting.
\item \textbf{Forecasting Horizon Sensitivity:} Measuring model performance across short- to long-term prediction intervals.
\item \textbf{Generalization to Unseen Sites:} Testing the ability to transfer across geographic regions with no prior exposure during training.
\item \textbf{Multivariate Output Forecasting:} Simultaneous prediction of multiple target variables, such as load and renewables generation.
\item \textbf{Probabilistic Forecasting:} Evaluating the quality and calibration of predicted uncertainty intervals.
\item \textbf{Context Window Sensitivity:}  Evaluation of the forecasting performance with access to varying look-back windows.
\end{itemize}

By unifying datasets, metrics and performance categories, we attempt to illustrate the instances when transformer and foundation models enhance the performance of power system forecasting, alongside their limitations. The resulting insights aim to guide both researchers exploring these models and practitioners considering their deployment in operational energy‑management systems.


The rest of the paper is organized as follows: Section 2 provides a detailed description of the dataset, whereas Section 3 presents the assessed models. Section 4 discusses the methodology for the evaluation of the forecasting models. Section 5 presents the data preparation pipeline and implementation details to ensure a fair comparison and reproducibility. Section 6 demonstrates the results of the conducted experiments, and Section 7 concludes the work and suggests areas for further study.
\section{Dataset Description}\label{sec:dataset}

This comparative study utilizes the publicly available dataset developed under the Advanced Research Projects Agency–Energy (ARPA-E) PERFORM initiative and released by the National Renewable Energy Laboratory (NREL) \cite{bryce2023solar, bryce2024solar}. This dataset contains a high-resolution, multi-variate, and geographically diverse collection of actual and forecasted time series, including load, solar and wind power generation across four major U.S. balancing authorities: the Midcontinent Independent System Operator (MISO), the New York Independent System Operator (NYISO), and the Southwest Power Pool (SPP) \cite{bryce2023solar}. The dataset spans two years, with both actuals and forecasts for 2017, and only actuals for 2018. For the purpose of computational efficiency, we focus
solely on ERCOT \cite{bryce2024solar}, which has high renewable shares that provide realistic yet manageable complexity \cite{ercotLoadProfiling}. Table \ref{ERCOT} provides a high-level overview of the ERCOT dataset used in this work.

\subsection{Generation of Actuals and Forecasts}
For the actuals, solar irradiance fields were collected from the NSRDB/Physical Solar Model and converted to AC power using the PVWatts/reV pipeline, leveraging plant-level metadata \cite{sengupta2018national}. Wind resource data were gathered from the WRF-based WIND Toolkit and was similarly converted to power \cite{draxl2015wind}. Zonal and system load profiles were obtained from ERCOT’s public five-minute archive \cite{ercotLoadProfiling}.
Deterministic forecasts were generated by down-scaling an ECMWF 51-member ensemble's weather variable forecasts, which then led to the same PVWatts/reV pipeline to produce day-ahead (DA) and intra-day (ID) power forecasts. An additional machine learning-based post-processor was applied to generate hour-ahead (HA) forecasts.

\begin{table}[!htbp]
\centering
\caption{Overview of the ERCOT Case Study Datasets}
\begin{tabular}{ccccc}
\hline
\textbf{Type} & \textbf{Existing Sites} & \textbf{Planned Sites} & \textbf{Total Capacity (GW)} & \textbf{Geographic Area}                                                   \\ \hline
Wind          & 125               & 139              & 21                     & \begin{tabular}[c]{@{}c@{}}site / zone / \\ system\end{tabular} \\
Solar         & 22                & 204              & 3                      & \begin{tabular}[c]{@{}c@{}}site / zone / \\ system\end{tabular} \\
Load          & N/A               & N/A              & 75 (peak)              & \begin{tabular}[c]{@{}c@{}}zone /\\ system\end{tabular}         \\ \hline
\end{tabular}
\label{ERCOT}
\end{table} 

For probabilistic forecasts, Bayesian Model Averaging \cite{doubleday2020probabilistic} was used to produce predictive distributions for solar and wind profiles, while an M$^{3}$ ensemble method was applied to load profiles, transforming the deterministic forecasts into full probabilistic outputs.

\subsection{Solar Component}
The solar component includes features such as 5-minute AC power, site coordinates, AC/DC ratings, inverter loading ratio, and commissioning year. Each profile contains not only power data but also meteorological variables, including global, direct, and clear-sky solar irradiance, 2-meter air temperature, surface pressure, and 10-meter wind components, all of which drive the actual and forecast generation models. Forecast datasets for solar consist of deterministic HA, ID, and DA time series profiles, along with BMA-derived $P01$–$P99$ quantiles.

\subsection{Wind Component}
The wind component is characterized by features such as 5-minute AC power, turbine location, hub height, rotor diameter, capacity factor, and commissioning year. Each profile includes 10-meter and 100-meter U/V wind components, surface pressure, and both the 2-meter temperature and dew-point data for each point. Forecast sets follow the same HA, ID, and DA horizon structure as solar but also include per-member deterministic profiles for all 51 ECMWF ensemble members.

\subsection{Load Component}
\begin{figure}[h]
    \centering
    \includegraphics[width=0.6\linewidth]{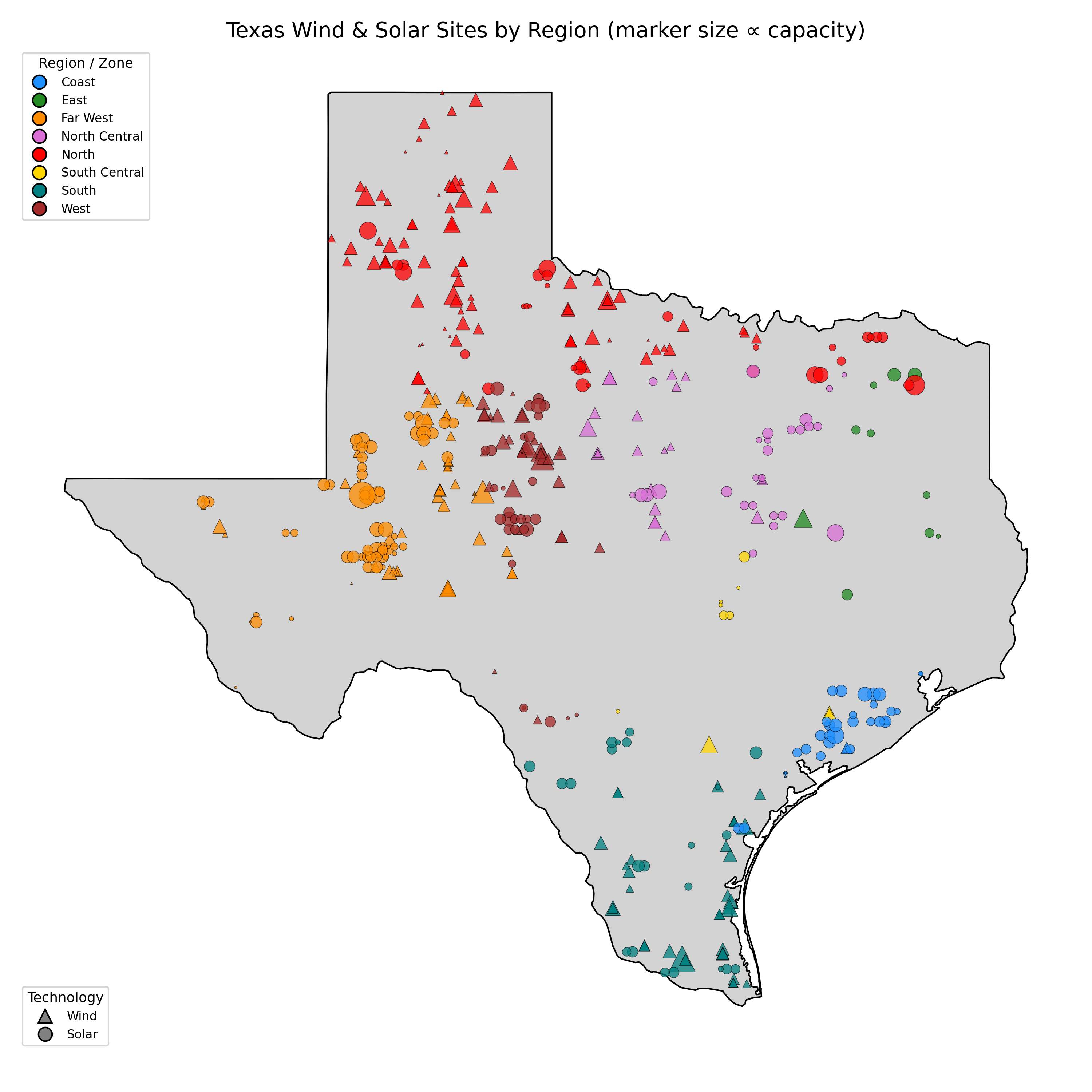}
    \caption{Texas ERCOT solar and wind sites relative to capacity}
    \label{fig:texas_1}
\end{figure}

The load component contains 5-minute load measurements for each of ERCOT’s eight load zones, as well as the system total. Profile contents include zone-level temperature, global horizontal irradiance (GHI), and dew-point temperature from three representative stations per zone, in addition to calendar indicators such as minute-of-day, day-of-week, month, and year. Forecast datasets for load are generated using a CNN-based deterministic model for HA, ID, and DA horizons, along with M$^{3}$-based probabilistic distributions. Figure \ref{fig:texas_1} provides a map of the locations of the individual wind and solar sites and an indication of their capacity, alongside the load regions.

\section{Forecasting Models}

This study evaluates a diverse set of forecasting models, categorized into two distinct families based on their training paradigm: Foundation Models (FMs) and Transformer Models, as well as classic deep learning methods as baselines. The primary distinction lies in their training data and methodology. FMs are large models pre-trained on massive, multi-domain time-series corpora (e.g., finance, weather, retail) and are subsequently applied to new tasks with zero or minimal fine-tuning. In contrast, transformer models in this work are moderately-sized transformer architectures trained entirely from scratch on the specific target dataset without any pre-trained checkpoints.

To ensure a fair and relevant comparison, all selected models are (i) trained mainly for time-series data (Large Language Models trained on textual data were excluded) (ii) open-source, (iii) publicly available pre-trained weights (for FMs) or official implementations for transformer models, and (iv) introduce a specific and unique implementation detail or addition that targets better forecasting performance.

\subsection{Foundation Models (FMs)}

This category includes large-scale models that leverage knowledge learned from vast datasets.

\begin{itemize}
\item \textbf{TimesFM} \cite{das2024decoder} is a decoder-only Transformer pre-trained on 100 billion time points. It processes time-series data by dividing it into patches, which are then passed through a causal self-attention mechanism. For probabilistic forecasting, its final hidden state parameterizes a mixture of Gaussian distributions, allowing for the derivation of any quantile. To enhance long-horizon accuracy efficiently, it employs a causal dilation schedule in its attention mechanism, creating a multi-scale receptive field that captures both local details and long-term context without causing quadratic computational costs.

\item \textbf{Chronos} \cite{ansari2024chronos} adapts a T5 \cite{raffel2020exploring} language model architecture for forecasting. It operates by tokenizing time-series values through affine scaling and then predicting future tokens using a standard cross-entropy objective. During inference, it samples token trajectories that are rescaled back into time-series values to form a predictive distribution. Its main contribution from the architecture perspective is the use of learned time-token offsets, where calendar features (like hour or day of the week) are embedded and added to value tokens. This allows the model to handle seasonality additively, freeing up model capacity.

\item \textbf{Moirai} \cite{liu2024moirai} is an encoder-only Transformer that uses a multi-patch projection technique, where segments of different lengths ($p_f$) are mapped to tokens using distinct learnable weights. It features an "any-variate" attention mechanism that flattens the time and variable axes, using Rotary Positional Embeddings (RoPE) \cite{su2024roformer} for time and a learned bias for variables. Its distinguishing feature is Cross-patch RoPE (RoPE-XS), which applies the same rotary angles to both time and variable dimensions, making it strong at identifying relationships between phase-shifted variables (e.g., weather patterns and load).

\item \textbf{MOMENT} \cite{goswami2024moment} is a Bidirectional Encoder Representations from Transformers (BERT)-based masked-patch encoder \cite{devlin2019bert} pre-trained on the "Time-series Pile" (1.1B points) \cite{goswami2024moment}. It reconstructs masked patches by minimizing a smooth Huber loss. A key aspect of its pre-training is a masked-patch approach, which begins by masking short, contiguous spans and progresses to masking larger, random blocks. This strategy forces the model to develop more global representations of the time series. It also utilizes Time2Vec positional encoding, which is designed to represent the monotonic trends often seen in most signals \cite{kazemi2019time2vec}.

\item \textbf{Tiny Time Mixer (TTM)} \cite{ekambaram2024tiny} is an lightweight model ($\sim$1M parameters) based on an Multi-Layer Perceptrons (MLP) Mixer architecture, making it suitable for deployment on resource-constrained devices. It processes data by alternating between mixing features along the temporal dimension ($g_t$) and the channel dimension ($f_c$). Despite its simplicity, it incorporates resolution-aware token shuffling to learn invariances and uses an expert-per-horizon gating mechanism, where different small expert networks specialize in short-term versus long-term forecasts without a significant increase in model size.
\end{itemize}

\subsection{Transformer Models}

In contrast to the FMs described above, this group consists of models trained exclusively on the target ERCOT dataset.

\begin{itemize}
    \item \textbf{Temporal Fusion Transformer (TFT)} \cite{lim2021temporal} is a comprehensive architecture designed to integrate static covariates, known future inputs, and historical time-varying data. It uses variable-selection networks to prune irrelevant inputs, an LSTM to encode local context, and multi-head attention to learn long-term dependencies. Its key feature is static context gating, where dedicated attention heads inject static metadata (like sensor location) into every temporal step of the forecast, improving generalization. It also enforces a monotone quantile loss to produce coherent, non-crossing prediction intervals.

\item \textbf{PatchTST} \cite{nie2022time} is a simple yet effective encoder-only transformer that achieves efficiency through channel independence. Each time-series variable is processed by a separate, weight-shared transformer, and the input series is first divided into non-overlapping patches. This reduces the attention mechanism's complexity from quadratic in the sequence length to quadratic in the number of patches. To maintain awareness of cross-variable dynamics, it uses a 2-D patch positional grid that encodes both temporal position and channel identity.

\item \textbf{TimeXer} \cite{wang2024timexer} is an exogenous-aware model that explicitly separates endogenous series from external variables. The architecture alternates between self-attention over patches of the endogenous series and cross-attention between the endogenous series and exogenous variable tokens. A set of global tokens acts as a bridge to channel information between them. Its main contribution is a causal cross-attention lag embedding, which allows the model to automatically learn the typical physical delays between external drivers (e.g., temperature) and the target series, improving lead-time alignment without manual feature engineering.
\end{itemize}

\subsection{Deep Learning Models}
To ground the performance of state-of-the-art models for energy forecasting from the literature, we include two widely-used deep learning architectures as baselines.

\begin{itemize}
    \item \textbf{Long Short-Term Memory (LSTM)} is a type of Recurrent Neural Network (RNN) specifically designed to capture long-range temporal dependencies. It processes data sequentially, using a series of gates (input, forget, and output) to regulate the flow of information through a cell state. This memory mechanism allows it to remember patterns over long periods, making it a standard choice for time-series forecasting. The final hidden state is fed into a dense layer to produce probabilistic quantile forecasts \cite{lecun2015deep}. It has been heavily applied in renewable forecasting and has demonstrated satisfactory performance \cite{wang2019review}.

    \item \textbf{Convolutional Neural Network (CNN)} while famous for image analysis, is highly effective for sequential data. A 1D-CNN applies convolutional filters across the input time series to automatically extract local, hierarchical features like trends, peaks, and seasonal shapes \cite{lecun2015deep}. By stacking layers of dilated convolutions, the model can efficiently increase its receptive field to capture dependencies over longer time scales. The extracted feature map is then flattened and passed to a prediction head for forecasting. It has also demonstrated effective performance in renewable and load forecasting \cite{aslam2021survey}.
\end{itemize}

Table \ref{tab:model_compare} summarizes the differences between the models described above. Params refers to the number of parameters in the model, whereas the context length points to the maximum number of tokens that the model can predict (i.e day-ahead forecasts at 5 minute resolution are considered 288 tokens). The pre-training points is the number of samples that the pre-trained checkpoint is trained on. The table also shows if the models natively support multivariate input/output and probabilistic forecasting.

\begin{table*}[!htbp]
 \caption{Key characteristics of the selected forecasting models}
 \label{tab:model_compare}
 \centering\small
 \begin{tabular}{lcccccc}
  \toprule
  \textbf{Model} & \textbf{Family} & \textbf{Params} & \textbf{Ctx. len.} &
  \textbf{Pre-train pts} & \textbf{Probabilistic} & \textbf{Multivariate} \\
  \midrule
  TimesFM        & FM         & 200M    & 512   & 100B     & Quantile heads (exp.) & Only Univariate \\
  Chronos-Bolt   & FM         & 710M    & 1024  & 84B      & Yes (Auto-regressive) & Only Univariate \\
  Moirai-L       & FM         & 311M    & 768   & 27B      & Yes (Mixture output)  & Multivariate \\
  MOMENT    & FM         & 125M    & 512   & 1.1B     & Yes (Mixture output)  & Multivariate \\
  TTM & FM   & 1M      & 256   & $\sim$1B & No                    & Multivariate \\
  \midrule
  TFT            & Transformer& $\sim$2M& 512   & —        & Yes (Quantile Loss)   & Multivariate \\
  PatchTST/64    & Transformer& 22M     & 512   & —        & No (can be extended)  & Multivariate \\
  TimeXer        & Transformer& 19M     & 512   & —        & Yes (Dropout/Ensemble)& Multivariate \\
  \bottomrule
 \end{tabular}
\end{table*}

\section{Methodology}

The primary objective of this study is to rigorously evaluate the performance and potential of transformer and foundation models for time-series forecasting in power system applications. Specifically, we aim to assess how these relatively large models perform on forecasting key operational grid-related signals, namely solar PV generation, wind power generation, and electric load, across a variety of forecasting settings. These models are benchmarked against task-specific deep learning forecasting baselines to determine the circumstances under which foundation models offer useful advantages or face critical limitations. The overarching goal is to understand whether the generalization and flexibility promised by foundation models can translate into improvements in practical forecasting workflows used by grid operators, forecasting vendors, and researchers.

\begin{figure}[t]
    \centering
    \includegraphics[width=0.9\linewidth, height=5.5cm]{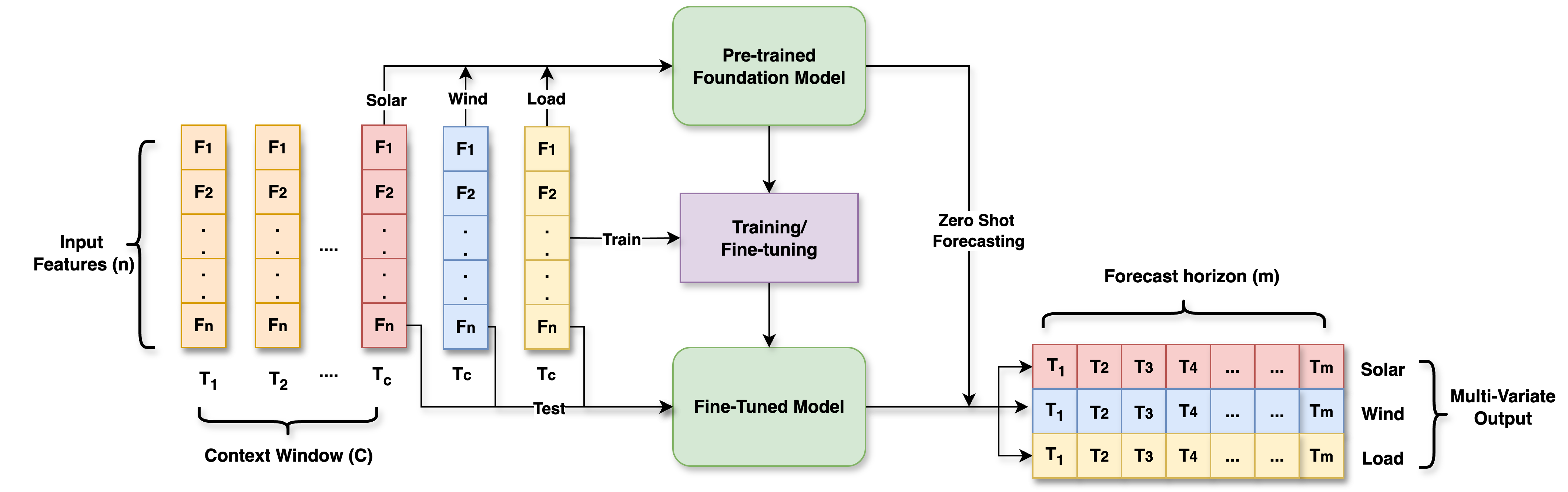}
    \caption{Illustration of Forecasting Attributes}
    \label{fig:methodology}
\end{figure}

To this end, we evaluate each model across eight complementary axes of performance, selected to highlight generalization, adaptability, multivariate input and output handling, and spatial generalization, which are core traits expected of foundation models. Figure \ref{fig:methodology} illustrates the data pipeline followed in this paper.

\subsection{Zero-Shot Forecasting}

This setting only applies to foundation models as there are pre-trained models available which are trained on large amounts of publicly available data. In the zero-shot setting, foundation models are evaluated directly on the 2018 time series data for solar, wind and load  without any additional supervised tuning or task-specific data. This scenario attempts to assess an important use case where foundation models could potentially generalize to grid-related applications without any labeled data. Performance is reported for solar, wind, and load signals across multiple ERCOT load zones and solar and wind sites. This experiment is conducted on day-ahead  forecasts only, with uni-variate input and output, and is limited only to foundation models.

\subsection{Fine-Tuning Capability}

We assess the ability of each model to improve with increasing amounts of task-specific data. Fine-tuning is performed using the 2017 data, incrementally increasing the training set in 20\% steps (i.e., 20\%, 40\%, ..., 100\%). This allows us to evaluate the data efficiency of the transformer and foundation models compared to deep learning algorithms. Each model is retrained or fine-tuned from a consistent starting point using the same optimization pipeline and hyperparameter values. This setting applies to all models and is conducted on day-ahead forecasting only, with uni-variate input and output.

\subsection{Forecasting Horizon Sensitivity}

Forecasting requirements in power systems span various time horizons depending on the application. Therefore, we evaluate the models' performance for three categories: short-term (15–60 minutes), intra-day (up to 6 hours), and day-ahead (24 hours). These horizons map to different operational tasks such as frequency regulation, economic dispatch, and energy market bidding. By evaluating the performance as a function of lead time, we attempt to identify whether some models are better suited for near-term or longer-term forecasting horizons. This experiment is conducted with full data training with uni-variate input and output.

\subsection{Generalization to Unseen Sites}

To measure spatial generalization, we evaluate model performance on sites that are not included in the training dataset. After full fine-tuning on a subset of locations (e.g., wind farms or load zones), we test the model on unseen sites within the same ISO (ERCOT). This experiment tests the extent to which learned representations can transfer to new sites with limited or non existent data, which is novel in this application domain, but could be very useful when new generation sites are coming online in rapid succession and to assess the potential of solar and wind generation in new sites. This experiment is conducted with full data training with uni-variate input and output and day-ahead forecasts only.

\subsection{Probabilistic Forecasting Performance}

We evaluate the probabilistic forecasting capabilities of models that support uncertainty quantification. For foundation models with quantile output heads or probabilistic decoding mechanisms, we compute the Continuous Ranked Probability Score (CRPS) to measure calibration and sharpness. 
Probabilistic forecasts provide essential situational awareness for system operators and possess potential for advancing decision making under uncertainty, particularly in grid reliability assessments and stochastic optimization.This experiment is conducted with full data training with uni-variate input and output and day-ahead forecasts only.

\subsection{Multivariate Input Handling}

To evaluate the models' ability to leverage multivariate temporal context, we compare univariate training, where the model receives only the target time series against multivariate training that includes additional input features such as temperature, irradiance, or wind speed. For models capable of multivariate input processing, we evaluate whether training with multi-dimensional input signals leads to improved forecasting accuracy or robustness. This experiment is conducted with full data training with day-ahead forecasts only and applies only to the models with the ability to handle multi-variate input.

\subsection{Multivariate Output Forecasting}

We also explore whether a single model performs better when it is jointly trained to predict solar, wind, and load values as a multivariate output, as opposed to training a separate model for each target. Since all the solar, wind sites and load zones are within the same ISO, 
the amount of utility scale solar generation also shows a natural correlation with demand, mainly driven by coinciding diurnal cycles and meteorological patterns, such as increased cooling loads during sunny periods.
Therefore, this setup tests the potential for shared feature extraction and cross-task learning, which may reduce model complexity and could potentially enhance forecasting accuracy. The multivariate output models are compared to to their single task counterparts. This experiment is conducted with full data training with day-ahead forecasts only.

\subsection{Context Window Sensitivity}

One strong feature of Foundation models and transformer-based architectures is their context window, which points to their ability to look back at recently observed values when forecasting future values. The context window limit often varies per model and architecture. We explore the difference in forecasting performance when varying the size of the context window (look-back duration). This experiment is conducted with full data training for day-ahead forecasts only and in the uni-variate setting.

\subsection{Evaluation Metrics}
 
We use the following three evaluation metrics:

\subsubsection{Normalized Mean Absolute Error (nMAE)}
The nMAE is computed as the average of the absolute forecast errors, normalized by the peak value of the actual time series:

\begin{equation}
    \text{nMAE} = \frac{1}{n} \sum_{t=1}^{n} \frac{|y_t - \hat{y}_t|}{y_{\text{max}}}
\end{equation}

where $y_t$ and $\hat{y}_t$ denote the actual and forecast values at time $t$, respectively, and $y_{\text{max}}$ is the maximum value of the target series.

\subsubsection{Normalized Root Mean Squared Error (nRMSE)}
The nRMSE penalizes larger deviations more heavily than nMAE and is similarly normalized by the maximum actual value:

\begin{equation}
    \text{nRMSE} = \sqrt{ \frac{1}{n} \sum_{t=1}^{n} \left( \frac{y_t - \hat{y}_t}{y_{\text{max}}} \right)^2 }
\end{equation}

\subsubsection{Continuous Ranked Probability Score (CRPS)}
The CRPS evaluates the accuracy of probabilistic forecasts by comparing the forecast cumulative distribution function $F$ against the observed values $x$:

\begin{equation}
    \text{CRPS}(F, x) = \int_{-\infty}^{\infty} \left( F(y) - \{ y \geq x \} \right)^2 \, dy
\end{equation}

Lower CRPS values indicate better-calibrated and sharper probabilistic forecasts. For models that generate multiple quantile forecasts, the CRPS is approximated via empirical integration over the predicted quantile levels.

\vspace{1em}

\section{Implementation}
This section describes the preparation and configuration of the data and experiments for evaluating the forecasting models. The aim is to ensure a consistent, reproducible, and fair comparison across all model families: foundation models, transformer models, and deep learning baselines.

\subsection{Data Preparation}
ERCOT data from the ARPA-E PERFORM dataset was used, spanning 2017--2018 at 5-minute resolution across multiple wind and solar sites and seven load zones. 12 months of data were allocated for training, 6 months for validation and hyper-parameter turning and 6 months for testing, with all of 2017 included in the training set and six non-consecutive months from 2018 in the evaluation and validation sets.

\subsection{Model Configuration and Training Protocol}
\textbf{Foundation Models:} Pre-trained checkpoints from official repositories were used when available; otherwise, implementations were sourced from open-source libraries such as \texttt{darts}. The context window was set to twice the forecast horizon where possible, otherwise truncated to the model’s maximum supported length.

\textbf{Transformer Models:} Implementations were drawn from official or trustworthy open-source repositories. Models were trained from scratch on the ERCOT dataset using identical optimization settings to ensure comparability.

\textbf{Deep Learning Baselines:} LSTM and 1D-CNN architectures were implemented in PyTorch, using comparable input/output structures, loss functions, and optimization settings to the transformer-based models. All models were trained using the Adam optimizer \cite{kingma2014adam} with early stopping based on validation performance. Learning rates, batch sizes, and other hyperparameters were tuned via Nixtla \cite{madhusudhanan2024hyperparameter}.

\subsection{Experimental Settings}
All experiments followed a standardized workflow to ensure comparability. Each model was trained and validated on the prepared training set, with hyperparameters fixed across experiments unless explicitly varied for a sensitivity study. Performance was evaluated on both seen and unseen subsets using the metrics described in Section~\ref{sec:methodology}. For fairness, data preprocessing, normalization, and batching procedures were identical across all model families. Missing values of the actual values were imputed using available forecasts when possible; otherwise, the dataset-provided imputation was used. Metrics (nMAE, nRMSE, CRPS) were computed per horizon and then averaged over the evaluation period for each task and dataset split. Random seeds were fixed for data splits, model initialization, and training to ensure reproducibility. 

All experiments were conducted on a server equipped with 2 A100 40GB GPUs, 8 intel (check type here) CPUs and 64 GB of RAM, running on the Linux operating system. Model training and evaluation were implemented in \texttt{Python~3.10}, using \texttt{Torch~2.8} and \texttt{CUDA~12.1}\cite{paszke2019pytorch}. Mixed-precision training was employed where supported to reduce memory usage and improve throughput.

\subsection{Seen/Unseen Configuration}
\begin{figure}[h]
    \centering
    \includegraphics[width=0.6\linewidth]{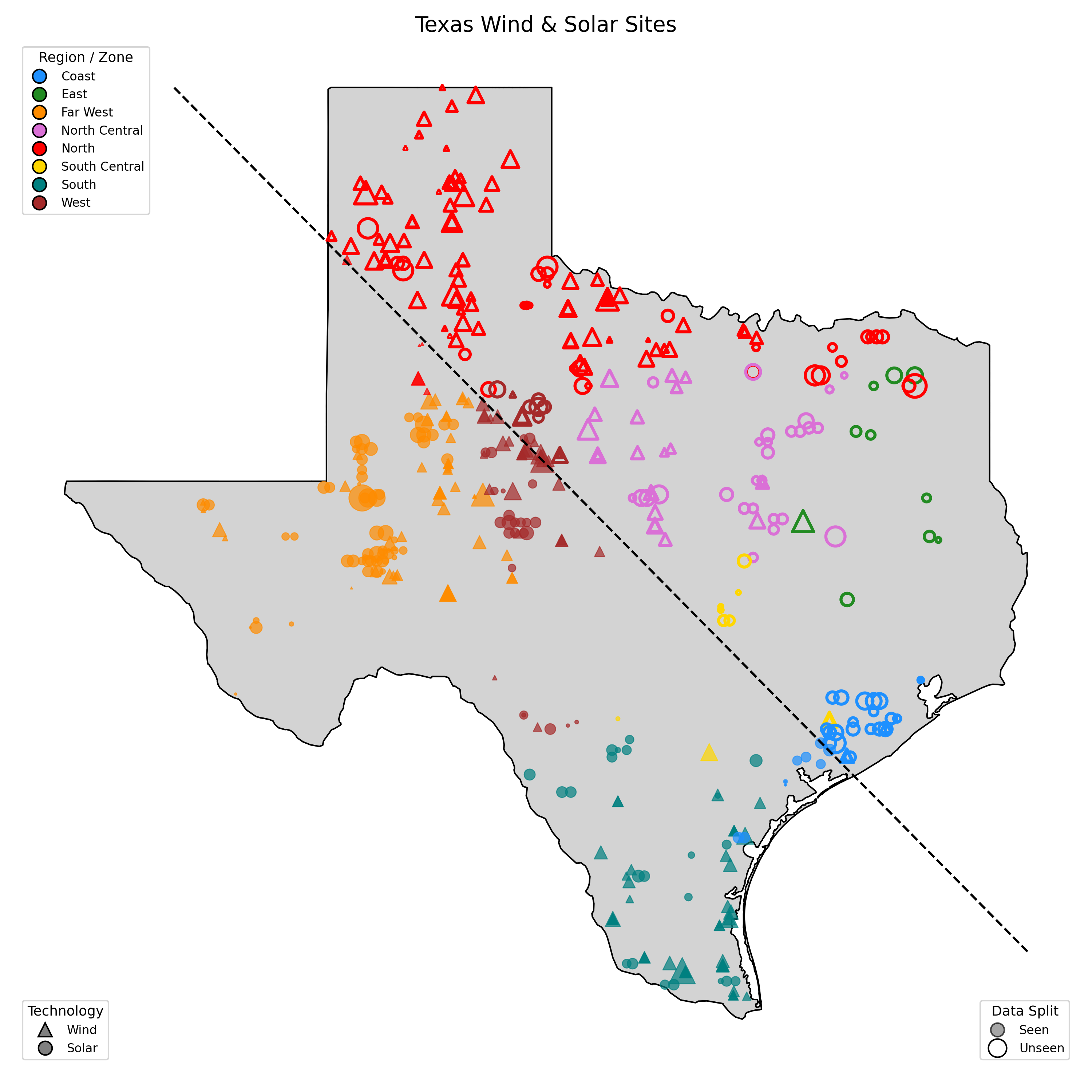}
    \caption{Geographical distribution of the seen (training) and unseen (testing) locations within the ERCOT system. A diagonal geographic boundary was established to test spatial generalization. Sites in the southwestern region (filled markers) were used for training, while approximately 45\% of utility-scale sites in the northeastern region (unfilled markers) were strictly withheld to evaluate the models' zero-shot forecasting capabilities on unseen sites.}
    \label{fig:texas_unseen}
\end{figure}

To evaluate spatial generalization capabilities, the dataset was partitioned into two distinct evaluation subsets:
\begin{itemize}
    \item \textbf{Seen sites:} Sites and zones that were included in the training dataset, used to assess in-domain forecasting performance.
    \item \textbf{Unseen sites:} Sites and zones withheld entirely from the training process to evaluate spatial generalization. As illustrated in Figure \ref{fig:texas_unseen}, a diagonal geographical split was applied to withhold approximately 45\% of the wind and solar sites located in the northeastern portion of the ERCOT grid.
\end{itemize}

Models were trained exclusively on the seen locations and then evaluated on both subsets.

\section{Results and Discussion}

This section presents the performance of foundation models, transformer models, and deep learning baselines across deterministic and probabilistic forecasting tasks. Throughout these experiments, solar and wind forecasting are evaluated on a per-site basis, while load forecasting is evaluated on a per-zone basis. Furthermore, all reported performance metrics are averaged across their respective sites or zones. The results correspond to the experiments described in Section 4

\subsection{Zero-Shot Forecasting}

\begin{table*}[ht]
\centering\small
\caption{Experiment 1 — Zero-shot deterministic nMAE / nRMSE (\%).}
\label{tab:exp1_zs_all}
\begin{tabular}{lcccccc}
\toprule
\multirow{2}{*}{\textbf{Foundation Model}} &
\multicolumn{2}{c}{\textbf{Solar}} &
\multicolumn{2}{c}{\textbf{Wind}} &
\multicolumn{2}{c}{\textbf{Load}} \\
\cmidrule(lr){2-3}\cmidrule(lr){4-5}\cmidrule(lr){6-7}
 & \textbf{MAE} & \textbf{RMSE} & \textbf{MAE} & \textbf{RMSE} & \textbf{MAE} & \textbf{RMSE} \\ \midrule
TimesFM   & 9.25 & 15.1 & 8.73 & 13.25 & 7.77 & 12.34 \\
Chronos-Bolt  & 11.21 & 17.43 & 11.31 & 18.17 & 9.18 & 15.08 \\
Moirai-L      & 10.91 & 17.5 & 12.1 & 18.71 & 8.89 & 14.22 \\
MOMENT  & 11.26 & 18.62 & 11.22 & 17.48 & 9.26 & 13.94 \\
TTM & 14.03 & 22.03 & 12.14 & 20.21 & 12.25 & 17.28 \\
\bottomrule
\end{tabular}
\end{table*}

Table \ref{tab:exp1_zs_all} shows that all evaluated foundation models exhibit relatively high errors in the zero-shot setting. While there is some variation between models, with TimesFM generally achieving the lowest errors, the overall performance levels are far from what would be required for operational use in solar, wind, or load forecasting. The magnitude of the nMAE and nRMSE values indicates that current foundation models, when applied directly without adaptation, struggle to capture the domain-specific dynamics of renewable generation and demand despite some of the model pre-training data including energy applications. The key takeaway from Table 3 is that while models like TimesFM demonstrate the lowest relative zero-shot errors, the absolute nMAE values (ranging from 7.7\% to over 14\%) remain far too high for practical grid operations. This underscores that out of the box foundation models are not yet capable of bypassing specialized training for power system forecasting tasks, indicating the need for further fine-tuning as explored in the subsequent sections.

\subsection{Fine-Tuning Capability}

\begin{table*}[h]
\caption{Experiment 2 - Solar — nMAE / nRMSE (\%) vs.\ training-set fraction (day-ahead, seen sites).}
\centering
\label{tab:ft_solar}
\begin{tabular}{lcccccccccc}
\hline
\multirow{2}{*}{\textbf{Model}} & \multicolumn{2}{c}{\textbf{20\%}} & \multicolumn{2}{c}{\textbf{40\%}} & \multicolumn{2}{c}{\textbf{60\%}} & \multicolumn{2}{c}{\textbf{80\%}} & \multicolumn{2}{c}{\textbf{100\%}} \\ \cline{2-11} 
                                & \textbf{MAE}    & \textbf{RMSE}   & \textbf{MAE}    & \textbf{RMSE}   & \textbf{MAE}    & \textbf{RMSE}   & \textbf{MAE}    & \textbf{RMSE}   & \textbf{MAE}    & \textbf{RMSE}    \\ \hline
TimesFM                & 8.36            & 13.82           & 8.23            & 12.04           & 7.73            & 10.08           & 5.30            & 8.73            & 3.71            & 7.02             \\
Chronos-Bolt           & 7.79            & 13.15           & 7.11            & 11.82           & 6.12            & 9.43            & 4.99            & 8.33            & 3.78            & 7.03             \\
Moirai-L               & 8.14            & 14.54           & 7.08            & 10.32           & 6.76            & 9.91            & 5.12            & 8.63            & 4.03            & 7.72             \\
MOMENT           & 7.98            & 12.45           & 6.95            & 9.86            & 6.22            & 9.14            & 5.31            & 8.54            & 4.18            & 8.05             \\
TTM        & 7.82            & 11.92           & 6.74            & 9.42            & 6.45            & 9.21            & 6.06            & 9.03            & 4.71            & 8.31             \\ \midrule
PatchTST               & 16.25           & 22.13           & 12.82           & 16.32           & 8.12            & 13.21           & 4.90            & 8.43            & 4.18            & 7.82             \\
TimeXer                & 14.15           & 18.2            & 12.61           & 16.52           & 9.11            & 13.44           & 5.03            & 8.12            & 3.85            & 5.82             \\
TFT                    & 16.34           & 21.01           & 13.62           & 18.47           & 8.53            & 12.24           & 4.88            & 7.58            & 3.91            & 6.02             \\ \midrule
LSTM                   & 27.13           & 33.14           & 22.33           & 27.91           & 15.23           & 21.11           & 9.15            & 14.05           & 6.71            & 9.04             \\
1-D CNN                & 22.45           & 29.4            & 18.14           & 23.32           & 12.42           & 17.41           & 8.87            & 12.94           & 6.34            & 8.85             \\ \hline
\end{tabular}
\end{table*}

\begin{table*}[h]
\caption{Experiment 2 - Wind — nMAE / nRMSE (\%) vs.\ training-set fraction (day-ahead, seen sites).}
\centering
\label{tab:ft_wind}
\begin{tabular}{lcccccccccc}
\hline
\multirow{2}{*}{\textbf{Model}} & \multicolumn{2}{c}{\textbf{\textbf{20\%}}} & \multicolumn{2}{c}{\textbf{40\%}} & \multicolumn{2}{c}{\textbf{60\%}} & \multicolumn{2}{c}{\textbf{80\%}} & \multicolumn{2}{c}{\textbf{100\%}} \\ \cline{2-11} 
                                & \textbf{MAE}    & \textbf{RMSE}   & \textbf{MAE}    & \textbf{RMSE}   & \textbf{MAE}    & \textbf{RMSE}   & \textbf{MAE}    & \textbf{RMSE}   & \textbf{MAE}    & \textbf{RMSE}    \\ \hline
TimesFM                & 8.51            & 14.05           & 8.35            & 12.26           & 7.89            & 10.34           & 5.45            & 8.91            & 3.84            & 7.18             \\
Chronos-Bolt           & 7.94            & 13.36           & 7.25            & 12.03           & 6.29            & 9.61            & 5.11            & 8.49            & 3.91            & 7.18             \\
Moirai-L               & 8.31            & 14.71           & 7.23            & 10.54           & 6.92            & 10.13           & 5.28            & 8.78            & 4.17            & 7.88             \\
MOMENT           & 8.12            & 12.63           & 7.09            & 10.08           & 6.41            & 9.35            & 5.45            & 8.69            & 4.33            & 8.22             \\
TTM         & 7.95            & 12.14           & 6.88            & 9.66            & 6.61            & 9.44            & 6.22            & 9.25            & 4.86            & 8.49             \\ \midrule
PatchTST               & 16.41           & 22.45           & 12.97           & 16.63           & 8.29            & 13.49           & 5.06            & 8.58            & 4.33            & 7.96             \\
TimeXer                & 14.31           & 18.45           & 12.76           & 16.77           & 9.29            & 13.62           & 5.18            & 8.26            & 4.01            & 7.96             \\
TFT                    & 16.51           & 21.24           & 13.78           & 18.71           & 8.69            & 12.47           & 5.02            & 7.73            & 4.08            & 7.19             \\ \midrule
LSTM                   & 27.36           & 33.39           & 22.54           & 28.12           & 15.41           & 21.35           & 9.28            & 14.21           & 6.85            & 9.23             \\
1-D CNN                & 22.66           & 29.63           & 18.36           & 23.58           & 12.58           & 17.62           & 9.01            & 13.14           & 6.51            & 9.03             \\ \hline
\end{tabular}
\end{table*}

\begin{table*}[h]
\caption{Experiment 2 - Load — nMAE / nRMSE (\%) vs.\ training-set fraction (day-ahead, seen zones).}
\centering
\label{tab:ft_load}
\begin{tabular}{lcccccccccc}
\hline
\multirow{2}{*}{\textbf{Model}} & \multicolumn{2}{c}{\textbf{\textbf{20\%}}} & \multicolumn{2}{c}{\textbf{40\%}} & \multicolumn{2}{c}{\textbf{60\%}} & \multicolumn{2}{c}{\textbf{80\%}} & \multicolumn{2}{c}{\textbf{100\%}} \\ \cline{2-11} 
                                & \textbf{MAE}    & \textbf{RMSE}   & \textbf{MAE}    & \textbf{RMSE}   & \textbf{MAE}    & \textbf{RMSE}   & \textbf{MAE}    & \textbf{RMSE}   & \textbf{MAE}    & \textbf{RMSE}    \\ \hline
TimesFM                & 5.11            & 7.24            & 4.28            & 6.03            & 3.75            & 5.32            & 3.10            & 4.53            & 2.82   & 4.05    \\
Chronos-Bolt          & 5.08            & 7.12            & 4.22            & 6.01            & 3.67            & 5.30            & 3.09            & 4.49            & 2.86            & 4.08             \\
Moirai-L              & 5.25            & 7.43            & 4.31            & 6.09            & 3.84            & 5.45            & 3.25            & 4.61            & 2.95            & 4.15             \\
MOMENT          & 5.19            & 7.18            & 4.37            & 6.14            & 3.80            & 5.40            & 3.22            & 4.59            & 3.01            & 4.19             \\
TTM        & 5.34            & 7.55            & 4.48            & 6.23            & 3.88            & 5.51            & 3.34            & 4.70            & 3.10            & 4.31             \\ \midrule
PatchTST              & 6.73            & 8.84            & 5.40            & 7.14            & 4.36            & 6.03            & 3.61            & 5.00            & 3.21            & 4.48             \\
TimeXer               & 6.42            & 8.55            & 5.19            & 6.93            & 4.22            & 5.92            & 3.48            & 4.87            & 3.09            & 4.36             \\
TFT                   & 6.81            & 8.94            & 5.55            & 7.31            & 4.39            & 6.08            & 3.59            & 4.98            & 3.18            & 4.42             \\ \midrule
LSTM                  & 9.15            & 11.11           & 7.32            & 8.89            & 5.65            & 7.12            & 4.30            & 5.64            & 3.76            & 5.02             \\
1-D CNN               & 8.40            & 10.35           & 6.88            & 8.40            & 5.22            & 6.75            & 4.05            & 5.34            & 3.62            & 4.85             \\ \hline
\end{tabular}
\end{table*}

Experiment 2 investigates how the amount of available training data influences forecasting performance, with results for solar, wind, and load presented in Tables \ref{tab:ft_solar}–\ref{tab:ft_load}. Across all variables, a clear and consistent trend emerges: increasing the fraction of the training set substantially reduces both nMAE and nRMSE values. Beyond the 80\% level, improvements become less noticeable, suggesting diminishing returns from additional data once the model has captured the dominant temporal patterns that occur within a one year timeframe.
These results collectively highlight that foundation models are highly data-efficient relative to conventional deep learning baselines, often achieving lower errors with only 20\% of the training data than LSTMs and CNNs achieve with the full 100\%. For exmaple, TimesFM and Chronos-Bolt consistently maintain nMAE values under 9\% for renewables. However, the continuous improvement observed up to the 80\% data mark proves that despite massive pre-training, these models still rely significantly on domain-specific fine-tuning to reach a satisfactory operational accuracy.

\subsection{Forecasting Horizon Sensitivity}

\begin{table*}[ht]
\centering\footnotesize
\setlength{\tabcolsep}{1.2pt}
\caption{Experiment 3 — nMAE / nRMSE (\%) at three look-ahead horizons on seen sites.}
\label{tab:exp3_hor_all}
\begin{tabular}{lccccccccccccccccccc}
\toprule
\multirow{3}{*}{\textbf{Model}} &
\multicolumn{6}{c}{\textbf{Solar}} &
\multicolumn{6}{c}{\textbf{Wind}} &
\multicolumn{6}{c}{\textbf{Load}} \\
\cmidrule(lr){2-7}\cmidrule(lr){8-13}\cmidrule(lr){14-19}
 & \multicolumn{2}{c}{\textbf{60 m}} & \multicolumn{2}{c}{\textbf{6 h}} & \multicolumn{2}{c}{\textbf{24 h}} &
   \multicolumn{2}{c}{\textbf{60 m}} & \multicolumn{2}{c}{\textbf{6 h}} & \multicolumn{2}{c}{\textbf{24 h}} &
   \multicolumn{2}{c}{\textbf{60 m}} & \multicolumn{2}{c}{\textbf{6 h}} & \multicolumn{2}{c}{\textbf{24 h}} \\
\cmidrule(lr){2-3}\cmidrule(lr){4-5}\cmidrule(lr){6-7}
\cmidrule(lr){8-9}\cmidrule(lr){10-11}\cmidrule(lr){12-13}
\cmidrule(lr){14-15}\cmidrule(lr){16-17}\cmidrule(lr){18-19}
 & \textbf{MAE} & \textbf{RMSE} & \textbf{MAE} & \textbf{RMSE} & \textbf{MAE} & \textbf{RMSE} &
   \textbf{MAE} & \textbf{RMSE} & \textbf{MAE} & \textbf{RMSE} & \textbf{MAE} & \textbf{RMSE} &
   \textbf{MAE} & \textbf{RMSE} & \textbf{MAE} & \textbf{RMSE} & \textbf{MAE} & \textbf{RMSE} \\ \midrule
TimesFM          & 2.13 & 3.34 & 2.58 & 3.98 & 3.71 & 7.02 & 1.82 & 2.93 & 2.43 & 3.72 & 3.84 & 7.18 & 1.01 & 1.65 & 1.23 & 1.91 & 2.82 & 4.05 \\
Chronos-Bolt     & 2.26 & 3.59 & 2.77 & 4.10 & 3.78 & 7.03 & 2.04 & 3.14 & 2.59 & 3.81 & 3.91 & 7.18 & 1.09 & 1.79 & 1.28 & 2.00 & 2.86 & 4.08 \\
Moirai-L         & 2.39 & 3.68 & 2.88 & 4.21 & 4.03 & 7.72 & 2.13 & 3.25 & 2.71 & 3.92 & 4.17 & 7.88 & 1.15 & 1.76 & 1.32 & 2.03 & 2.95 & 4.15 \\
MOMENT      & 2.47 & 3.83 & 2.96 & 4.33 & 4.18 & 8.05 & 2.21 & 3.34 & 2.77 & 4.04 & 4.33 & 8.22 & 1.18 & 1.81 & 1.38 & 2.09 & 3.01 & 4.19 \\
TTM   & 2.74 & 4.06 & 3.29 & 4.61 & 4.71 & 8.31 & 2.45 & 3.62 & 3.09 & 4.31 & 4.86 & 8.49 & 1.32 & 2.01 & 1.47 & 2.27 & 3.10 & 4.31 \\ \midrule
PatchTST      & 2.62 & 3.89 & 3.11 & 4.42 & 4.18 & 7.82 & 2.36 & 3.51 & 2.96 & 4.18 & 4.33 & 7.96 & 1.28 & 1.94 & 1.42 & 2.17 & 3.21 & 4.48 \\
TimeXer          & 2.65 & 3.92 & 3.13 & 4.45 & 3.85 & 5.82 & 2.32 & 3.48 & 2.98 & 4.22 & 4.01 & 5.96 & 1.30 & 1.92 & 1.43 & 2.21 & 3.09 & 4.36 \\
TFT              & 2.71 & 3.97 & 3.18 & 4.49 & 3.91 & 6.02 & 2.39 & 3.60 & 3.08 & 4.28 & 4.08 & 6.19 & 1.34 & 1.96 & 1.45 & 2.23 & 3.18 & 4.42 \\ \midrule
LSTM             & 3.12 & 4.58 & 3.77 & 5.32 & 6.71 & 9.04 & 2.84 & 4.11 & 3.59 & 4.97 & 6.85 & 9.23 & 1.48 & 2.33 & 1.70 & 2.59 & 3.76 & 5.02 \\
1-D CNN          & 3.19 & 4.72 & 3.88 & 5.41 & 6.34 & 8.85 & 2.91 & 4.20 & 3.67 & 5.09 & 6.51 & 9.03 & 1.58 & 2.44 & 1.81 & 2.72 & 3.62 & 4.85 \\
\bottomrule
\end{tabular}
\end{table*}

Experiment 3 examines model performance across three forecasting horizons—short-term (60 minutes), medium-term (6 hours), and day-ahead (24 hours)—for solar, wind, and load time series. Across all targets, errors increase systematically with horizon length, reflecting the accumulation of uncertainty as forecasts extend further into the future. Overall, these results highlight two important patterns:
(i) Forecasting skill decays predictably with horizon length across all variables and model types.
(ii) Foundation models deliver the largest relative advantage in short- and medium-term horizons, particularly for renewable energy forecasting. As seen in Table 7, top foundation models achieve 60-minute ahead nMAE scores nearly half those of the deep learning baselines for solar and wind. However, as the horizon extends to 24 hours, this performance gap noticeably narrows, indicating that traditional architectures become relatively more competitive as long-term uncertainty dominates.

\subsection{Generalization to Unseen Sites}

\begin{table*}[h]
\centering\small
\caption{Experiment 4 — nMAE and nRMSE (\%) on \emph{seen} and \emph{unseen} sites (day-ahead).}
\begin{tabular}{lcccccccccccc}
\hline
\multirow{3}{*}{\textbf{Model}} 
& \multicolumn{4}{c}{\textbf{Solar}} 
& \multicolumn{4}{c}{\textbf{Wind}} 
& \multicolumn{4}{c}{\textbf{Load}} \\ \cline{2-13} 
& \multicolumn{2}{c}{\textbf{Seen}} & \multicolumn{2}{c}{\textbf{Unseen}} 
& \multicolumn{2}{c}{\textbf{Seen}} & \multicolumn{2}{c}{\textbf{Unseen}} 
& \multicolumn{2}{c}{\textbf{Seen}} & \multicolumn{2}{c}{\textbf{Unseen}} \\ \cline{2-13} 
& \textbf{MAE} & \textbf{RMSE} & \textbf{MAE} & \textbf{RMSE} 
& \textbf{MAE} & \textbf{RMSE} & \textbf{MAE} & \textbf{RMSE} 
& \textbf{MAE} & \textbf{RMSE} & \textbf{MAE} & \textbf{RMSE} \\ \hline
TimesFM          & 3.71 & 7.02 & 4.73 & 8.03 & 3.84 & 7.18 & 4.55 & 8.37 & 2.82 & 4.05 & 3.11 & 4.02 \\
Chronos-Bolt     & 3.78 & 7.03 & 4.91 & 8.01 & 3.91 & 7.18 & 4.83 & 9.12 & 2.86 & 4.08 & 3.03 & 4.05 \\
Moirai-L         & 4.03 & 7.72 & 5.28 & 9.11 & 4.17 & 7.88 & 5.11 & 9.14 & 2.95 & 4.15 & 3.99 & 5.72 \\
MOMENT           & 4.18 & 8.05 & 4.86 & 8.31 & 4.33 & 8.22 & 5.05 & 8.89 & 3.01 & 4.19 & 3.28 & 4.57 \\
TTM              & 4.71 & 8.31 & 6.21 & 9.07 & 4.86 & 8.49 & 6.83 & 9.86 & 3.10 & 4.31 & 5.13 & 8.11 \\ \midrule
PatchTST         & 4.18 & 7.82 & 6.76 & 9.45 & 4.33 & 7.96 & 7.11 & 10.10 & 3.21 & 4.48 & 5.22 & 8.45 \\
TimeXer          & 3.85 & 5.82 & 6.81 & 9.52 & 4.01 & 5.96 & 6.53 & 8.77 & 3.09 & 4.36 & 5.64 & 8.53 \\ 
TFT              & 3.91 & 6.02 & 6.11 & 9.08 & 4.08 & 6.19 & 5.32 & 8.12 & 3.18 & 4.42 & 3.71 & 4.92 \\ \midrule
LSTM             & 6.71 & 9.04 & 12.35 & 15.84 & 6.85 & 9.23 & 11.42 & 18.33 & 3.76 & 5.02 & 6.24 & 8.21 \\
1-D CNN          & 6.34 & 8.85 & 11.49 & 14.97 & 6.51 & 9.03 & 12.20 & 18.94 & 3.62 & 4.85 & 6.04 & 7.83 \\ \hline
\end{tabular}
\end{table*}

Experiment 4 compares model performance between seen and unseen sites for day-ahead forecasts across solar, wind, and load. The results in Table \ref{tab:exp4_seen_unseen} reveal a clear generalization gap: all models experience higher errors when applied to unseen locations, reflecting the site-specific nature of renewable generation patterns and, to a lesser extent, load profiles. For solar, foundation models such as TimesFM and Chronos-Bolt exhibit the smallest increases in nMAE when moving from seen to unseen sites, suggesting that large-scale pretraining may provide some degree of spatial transferability. Nonetheless, even the best-performing models incur around a 1\% absolute nMAE increase, indicating that domain adaptation or localized fine-tuning would be beneficial for operational accuracy. In the wind domain, the generalization gap is more evident. Wind speed and direction are strongly influenced by topographical variations, making unseen site prediction inherently more challenging. Here, error increases are typically in the range of 0.7–2.0\% nMAE, with foundation models again outperforming deep learning baselines but still facing substantial degradation. Load forecasting shows the smallest seen–unseen gap, likely due to the smoother and more predictable nature of demand patterns compared to renewables.

Overall, Experiment 4 highlights that while foundation models provide better cross-site generalization than conventional baselines. 
These findings reinforce the necessity of localized fine-tuning when deploying models to new locations. Specifically, Table 8 demonstrates that while foundation models suffer a 1-2\% nMAE penalty on unseen sites, they still drastically outperform baselines like LSTM, which see their errors nearly double. Therefore, if spatial zero-shot transfer is strictly required—such as for newly commissioned sites lacking historical telemetry—foundation models offer the most robust fallback performance.

\subsection{Probabilistic Forecasting Performance}

\begin{table}[ht]
\centering\small
\caption{Experiment 5 — CRPS (\% of nominal) on seen sites (probabilistic-capable models) - Day ahead.}
\label{tab:exp5_crps}
\begin{tabular}{lccc}
\toprule
\textbf{Model} & \textbf{Solar} & \textbf{Wind} & \textbf{Load} \\ \midrule
TimesFM         & 2.53 & 2.13 & 1.74 \\
Chronos-Bolt    & 2.25          & 2.02          & 1.72 \\
Moirai-L        & 3.38          & 3.14          & 2.23 \\
MOMENT     & 2.87          & 2.92          & 1.92 \\ \midrule
TimeXer              & 2.84          & 2.37          & 1.81 \\
TFT                  & 3.18          & 2.30          & 1.76 \\ \midrule
LSTM                 & 3.32          & 3.14          & 2.11 \\
1-D CNN              & 3.28          & 3.51          & 2.86 \\
\bottomrule
\end{tabular}
\end{table}

Experiment 5 evaluates the probabilistic forecasting skill of models using the Continuous Ranked Probability Score (CRPS) for day-ahead forecasts on seen sites. Lower CRPS values indicate sharper and better-calibrated predictive distributions. These results demonstrate that large-scale pretraining significantly improves the reliability of probabilistic outputs compared to traditional baselines. Specifically, Table 9 shows Chronos-Bolt achieving the best overall calibration across all domains, with CRPS values as low as 1.72\% for load and 2.02\% for wind. This indicates that autoregressive and mixture-output foundation models are highly capable of rigorous uncertainty quantification, which is a critical requirement for stochastic grid optimization tasks.

\begin{figure}[h]
    \centering
    \includegraphics[width=0.6\linewidth]{figs/reliability_diagram_all.png}
    \caption{Reliability diagram for day-ahead probabilistic forecasts. The foundation and transformer models tightly track the ideal calibration line, while the deep learning model show overconfidence (S-curve) and poor calibration.}
    \label{fig:chronos}
\end{figure}

The P-P plots (Figure 4) clearly illustrates the differences in probabilistic calibration across the three model families. While the deep learning baseline shows an S-curve characteristic of overconfident and poorly calibrated forecasts, the transformer and foundation models progressively align with the ideal diagonal. Specifically, the Chronos foundation model achieves superior calibration, demonstrating its superior capacity to provide the reliable uncertainty bounds which are necessary for risk-aware grid optimization

\subsection{Multivariate Input Handling}

\begin{table*}[ht]
\centering\footnotesize
\setlength{\tabcolsep}{3.4pt}
\caption{Experiment 6 — Effect of adding weather covariates: nMAE / nRMSE (\%) on seen sites (day-ahead).}
\label{tab:exp6_mv_input}
\begin{tabular}{lcccccccccccc}
\toprule
\multirow{3}{*}{\textbf{Model}} &
\multicolumn{4}{c}{\textbf{Solar}} &
\multicolumn{4}{c}{\textbf{Wind}}  &
\multicolumn{4}{c}{\textbf{Load}}  \\
\cmidrule(lr){2-5}\cmidrule(lr){6-9}\cmidrule(lr){10-13}
 & \multicolumn{2}{c}{\textbf{UNI}} & \multicolumn{2}{c}{\textbf{+W}}%
 & \multicolumn{2}{c}{\textbf{UNI}} & \multicolumn{2}{c}{\textbf{+W}}%
 & \multicolumn{2}{c}{\textbf{UNI}} & \multicolumn{2}{c}{\textbf{+W}} \\
\cmidrule(lr){2-3}\cmidrule(lr){4-5}\cmidrule(lr){6-7}\cmidrule(lr){8-9}\cmidrule(lr){10-11}\cmidrule(lr){12-13}
 & \textbf{MAE} & \textbf{RMSE} & \textbf{MAE} & \textbf{RMSE} &
   \textbf{MAE} & \textbf{RMSE} & \textbf{MAE} & \textbf{RMSE} &
   \textbf{MAE} & \textbf{RMSE} & \textbf{MAE} & \textbf{RMSE} \\ \midrule
Moirai-L        & 4.03 & 7.72 & 3.65 & 7.10 & 4.17 & 7.88 & 3.78 & 7.20 & 2.95 & 4.15 & 2.74 & 3.95 \\
MOMENT-Base     & 4.18 & 8.05 & 3.79 & 7.40 & 4.33 & 8.22 & 3.98 & 7.50 & 3.01 & 4.19 & 2.83 & 4.00 \\
PatchTST/64     & 4.18 & 7.82 & 3.74 & 7.18 & 4.33 & 7.96 & 3.88 & 7.31 & 3.21 & 4.48 & 3.02 & 4.26 \\ \midrule
TimeXer         & 3.85 & 5.82 & 3.53 & 5.40 & 4.01 & 5.96 & 3.65 & 5.70 & 3.09 & 4.36 & 2.95 & 4.12 \\
TFT             & 3.91 & 6.02 & 3.66 & 5.78 & 4.08 & 6.19 & 3.71 & 6.00 & 3.18 & 4.42 & 3.01 & 4.28 \\ \midrule
LSTM            & 6.71 & 9.04 & 6.70 & 9.00 & 6.85 & 9.23 & 6.84 & 9.20 & 3.76 & 5.02 & 3.75 & 5.00 \\
1-D CNN         & 6.34 & 8.85 & 6.35 & 8.86 & 6.51 & 9.03 & 6.50 & 9.02 & 3.62 & 4.85 & 3.61 & 4.84 \\
\bottomrule
\end{tabular}
\end{table*}

Experiment 6 examines the impact of incorporating weather covariates (+W) into model inputs compared to univariate (UNI) historical target data alone. The evaluation is performed on seen sites for day-ahead forecasts of solar, wind, and load. For solar, the addition of weather features leads to consistent and noticeable performance gains across all foundation models. For instance, Moirai-L’s nMAE drops from 4.03\% to 3.65\%, while PatchTST improves from 4.18\% to 3.74\%. These improvements confirm the high predictive value of exogenous weather variables (e.g., irradiance forecasts, cloud cover) in capturing upcoming generation fluctuations that historical production alone cannot anticipate. In wind forecasting, the benefit of adding weather covariates is similarly clear, with reductions in both nMAE and nRMSE across nearly all models. Given the inherently weather-driven nature of wind generation, these results underline the necessity of incorporating meteorological forecasts into operational models, especially for horizons where persistence-based trends are less informative. Load forecasting also benefits from weather covariates, but the magnitude of improvement is smaller—typically in the range of 0.1–0.3\% nMAE reduction. This aligns with the fact that while load is partially weather-dependent (temperature, humidity), socio-economic and behavioral patterns also play significant roles.

Interestingly, classical deep learning baselines such as LSTM and 1-D CNN show minimal to no improvement from adding weather covariates. In these traditional architectures, exogenous features are typically concatenated with historical target values at each time step. For LSTMs, this forces the model to compress all multivariate interactions into a single, sequentially updated hidden state, creating an information bottleneck that dilutes the impact of specific weather events. Similarly, 1-D CNNs rely on fixed receptive fields that struggle to dynamically isolate cross-variable dependencies. This performance gap strongly highlights the effectiveness of attention mechanisms. By utilizing self-attention and cross-attention, modern transformer architectures can dynamically assign unique weights to specific weather variables at any given time step, allowing them to explicitly link a meteorological driver—such as an upcoming drop in irradiance—directly to the target generation without the constraints of sequential degradation

\subsection{Multivariate Output Forecasting}

\begin{table*}[ht]
\centering\footnotesize
\setlength{\tabcolsep}{3.2pt}
\caption{Experiment 7 - Day-ahead deterministic nMAE / nRMSE (\%) for \emph{single-task} (ST) vs.\ \emph{joint S+W+L} (JT) fine-tunes on seen sites.}
\label{tab:mv_out_all}
\begin{tabular}{lccccccccccccccc}
\toprule
\multirow{3}{*}{\textbf{Model}} &
\multicolumn{4}{c}{\textbf{Solar}} &
\multicolumn{4}{c}{\textbf{Wind}} &
\multicolumn{4}{c}{\textbf{Load}} \\
\cmidrule(lr){2-5}\cmidrule(lr){6-9}\cmidrule(lr){10-13}
 & \multicolumn{2}{c}{\textbf{MAE}} & \multicolumn{2}{c}{\textbf{RMSE}}%
 & \multicolumn{2}{c}{\textbf{MAE}} & \multicolumn{2}{c}{\textbf{RMSE}}%
 & \multicolumn{2}{c}{\textbf{MAE}} & \multicolumn{2}{c}{\textbf{RMSE}} \\
\cmidrule(lr){2-3}\cmidrule(lr){4-5}\cmidrule(lr){6-7}\cmidrule(lr){8-9}\cmidrule(lr){10-11}\cmidrule(lr){12-13}
 & \textbf{ST} & \textbf{JT} & \textbf{ST} & \textbf{JT} & \textbf{ST} & \textbf{JT} & \textbf{ST} & \textbf{JT} & \textbf{ST} & \textbf{JT} & \textbf{ST} & \textbf{JT} \\ \midrule
Moirai-L         & 4.03 & 3.92 & 7.72 & 7.58 & 4.17 & 4.21 & 7.88 & 7.95 & 2.95 & 2.73 & 4.15 & 3.91 \\
MOMENT      & 4.18 & 4.22 & 8.05 & 8.12 & 4.33 & 4.25 & 8.22 & 8.09 & 3.01 & 2.80 & 4.19 & 3.95 \\
TTM   & 4.71 & 4.75 & 8.31 & 8.39 & 4.86 & 4.94 & 8.49 & 8.60 & 3.10 & 2.89 & 4.31 & 4.07 \\ \midrule
PatchTST     & 4.18 & 4.26 & 7.82 & 7.94 & 4.33 & 4.38 & 7.96 & 8.03 & 3.21 & 3.01 & 4.48 & 4.22 \\
TimeXer          & 3.85 & 3.90 & 5.82 & 5.88 & 4.01 & 4.08 & 5.96 & 6.02 & 3.09 & 2.89 & 4.36 & 4.10 \\
TFT              & 3.91 & 3.84 & 6.02 & 5.92 & 4.08 & 4.01 & 6.19 & 6.09 & 3.18 & 2.95 & 4.42 & 4.16 \\ \midrule
LSTM             & 6.71 & 6.73 & 9.04 & 9.08 & 6.85 & 6.88 & 9.23 & 9.25 & 3.76 & 3.74 & 5.02 & 5.00 \\
1-D CNN          & 6.34 & 6.35 & 8.85 & 8.88 & 6.51 & 6.54 & 9.03 & 9.06 & 3.62 & 3.61 & 4.85 & 4.83 \\
\bottomrule
\end{tabular}
\end{table*}

Experiment 7 evaluates the effect of joint multi-task training across solar, wind, and load (\textbf{JT}) compared to individual single-task (\textbf{ST}) fine-tuning. Results are presented for day-ahead forecasts on seen sites. For solar forecasting, multi-task learning provides moderate improvements for most foundation models. Moirai-L sees a reduction from 4.03\% to 3.92\% in nMAE, and TFT improves from 3.91\% to 3.84\%. However, the gains are relatively small, suggesting that while some shared representations between domains can be beneficial, the benefits for solar are not as strong as those from adding domain-specific weather covariates (Experiment 6). For wind, joint training has a more mixed outcome. Some models such as TFT and TimeXer experience slight reductions in error, while others like Moirai-L and MOMENT show negligible or even small degradations. This suggests that, for wind forecasting, the cross-task transfer from solar and load may introduce domain noise that offsets potential benefits.For load, the improvements from joint training are more consistent. Across nearly all models, JT configurations outperform ST. For example, Moirai-L’s nMAE drops from 2.95\% to 2.73\%, and TimeXer improves from 3.09\% to 2.89\%. This is likely because load profiles are influenced by a combination of weather and generation patterns, which make them more receptive to cross-domain feature learning. Classical baselines such as LSTM and 1-D CNN show almost no difference between ST and JT, which reinforces that that transformer and attention based architectures are more capable of exploiting cross-task correlations.

\subsection{Context Window Length}

\begin{table*}[ht]
\centering\footnotesize
\setlength{\tabcolsep}{1.2pt}
\caption{Experiment 8 — Effect of context window length on day-ahead deterministic nMAE / nRMSE (\%) for \emph{seen} sites.}
\begin{tabular}{lccccccccccccccccccc}
\toprule
\multirow{3}{*}{\textbf{Model}} &
\multicolumn{6}{c}{\textbf{Solar}} &
\multicolumn{6}{c}{\textbf{Wind}} &
\multicolumn{6}{c}{\textbf{Load}} \\
\cmidrule(lr){2-7}\cmidrule(lr){8-13}\cmidrule(lr){14-19}
 & \multicolumn{2}{c}{\textbf{60 m}} & \multicolumn{2}{c}{\textbf{6 h}} & \multicolumn{2}{c}{\textbf{24 h}} &
   \multicolumn{2}{c}{\textbf{60 m}} & \multicolumn{2}{c}{\textbf{6 h}} & \multicolumn{2}{c}{\textbf{24 h}} &
   \multicolumn{2}{c}{\textbf{60 m}} & \multicolumn{2}{c}{\textbf{6 h}} & \multicolumn{2}{c}{\textbf{24 h}} \\
\cmidrule(lr){2-3}\cmidrule(lr){4-5}\cmidrule(lr){6-7}
\cmidrule(lr){8-9}\cmidrule(lr){10-11}\cmidrule(lr){12-13}
\cmidrule(lr){14-15}\cmidrule(lr){16-17}\cmidrule(lr){18-19}
 & \textbf{MAE} & \textbf{RMSE} & \textbf{MAE} & \textbf{RMSE} & \textbf{MAE} & \textbf{RMSE} &
   \textbf{MAE} & \textbf{RMSE} & \textbf{MAE} & \textbf{RMSE} & \textbf{MAE} & \textbf{RMSE} &
   \textbf{MAE} & \textbf{RMSE} & \textbf{MAE} & \textbf{RMSE} & \textbf{MAE} & \textbf{RMSE} \\ \midrule
TimesFM       & 4.28 & 7.62 & 3.95 & 7.15 & 3.71 & 7.02 & 4.45 & 7.82 & 4.05 & 7.35 & 3.84 & 7.18 & 3.05 & 4.38 & 2.90 & 4.15 & 2.82 & 4.05 \\
Chronos-Bolt  & 4.36 & 7.70 & 4.02 & 7.22 & 3.78 & 7.03 & 4.53 & 7.90 & 4.12 & 7.42 & 3.91 & 7.18 & 3.09 & 4.41 & 2.93 & 4.18 & 2.86 & 4.08 \\
Moirai-L      & 4.58 & 8.05 & 4.20 & 7.60 & 4.03 & 7.72 & 4.76 & 8.28 & 4.33 & 7.75 & 4.17 & 7.88 & 3.18 & 4.50 & 3.05 & 4.32 & 2.95 & 4.15 \\
MOMENT        & 4.65 & 8.18 & 4.30 & 7.72 & 4.18 & 8.05 & 4.84 & 8.40 & 4.41 & 7.88 & 4.33 & 8.22 & 3.22 & 4.53 & 3.10 & 4.35 & 3.01 & 4.19 \\
TTM           & 5.12 & 8.65 & 4.90 & 8.42 & 4.71 & 8.31 & 5.28 & 8.85 & 5.05 & 8.61 & 4.86 & 8.49 & 3.35 & 4.68 & 3.20 & 4.50 & 3.10 & 4.31 \\ \midrule
PatchTST      & 4.85 & 8.32 & 4.45 & 7.90 & 4.18 & 7.82 & 5.02 & 8.55 & 4.61 & 8.12 & 4.33 & 7.96 & 3.28 & 4.55 & 3.14 & 4.40 & 3.21 & 4.48 \\
TimeXer       & 4.50 & 6.45 & 4.12 & 6.08 & 3.85 & 5.82 & 4.65 & 6.60 & 4.28 & 6.22 & 4.01 & 5.96 & 3.15 & 4.42 & 3.12 & 4.38 & 3.09 & 4.36 \\
TFT           & 4.55 & 6.60 & 4.20 & 6.22 & 3.91 & 6.02 & 4.70 & 6.75 & 4.33 & 6.35 & 4.08 & 6.19 & 3.24 & 4.48 & 3.20 & 4.43 & 3.18 & 4.42 \\ \midrule
LSTM          & 7.10 & 9.45 & 6.88 & 9.22 & 6.71 & 9.04 & 7.25 & 9.60 & 7.02 & 9.38 & 6.85 & 9.23 & 3.82 & 5.08 & 3.79 & 5.05 & 3.76 & 5.02 \\
1-D CNN       & 6.75 & 9.20 & 6.50 & 8.95 & 6.34 & 8.85 & 6.90 & 9.35 & 6.65 & 9.12 & 6.51 & 9.03 & 3.68 & 4.92 & 3.65 & 4.89 & 3.62 & 4.85 \\ \hline
\end{tabular}
\end{table*}

Experiment 8 examines the impact of context window length—the amount of historical data available to the model—on day-ahead deterministic forecasting for seen sites. We compare short (6-hour), medium (12-hour), and long (24-hour) look-back periods across solar, wind, and load. For solar forecasting, longer context windows consistently improve performance across all model classes. TimesFM’s nMAE improves from 4.28\% at 6h to 3.71\% at 24h, and similar gains are seen for Chronos-Bolt (4.36\% $\rightarrow$ 3.78\%) and Moirai-L (4.58\% $\rightarrow$ 4.03\%). This trend reflects the value of including diurnal patterns and more complete weather evolution cycles in the model’s input. Wind forecasting shows a similar but slightly less pronounced improvement with longer context. For example, TimeXer improves from 4.65\% to 4.01\% nMAE, while Moirai-L drops from 4.76\% to 4.17\%. This smaller improvement margin may stem from the inherently higher variability and less predictable short-term fluctuations of wind compared to solar. In load forecasting the effect of increasing the context window is minimal beyond 12 hours. TimesFM’s nMAE improves modestly from 3.05\% (12h) to 2.82\% (24h), and similar small changes are observed for most other models. Classical baselines (LSTM and 1-D CNN) show the same general patterns but with smaller relative gains. For instance, LSTM’s solar nMAE improves from 7.10\% to 6.71\%, indicating that while longer input windows are beneficial, the architectural capacity of these models may limit their ability to fully exploit additional historical context.

\begin{figure}
    \centering
    \includegraphics[width=0.5\linewidth]{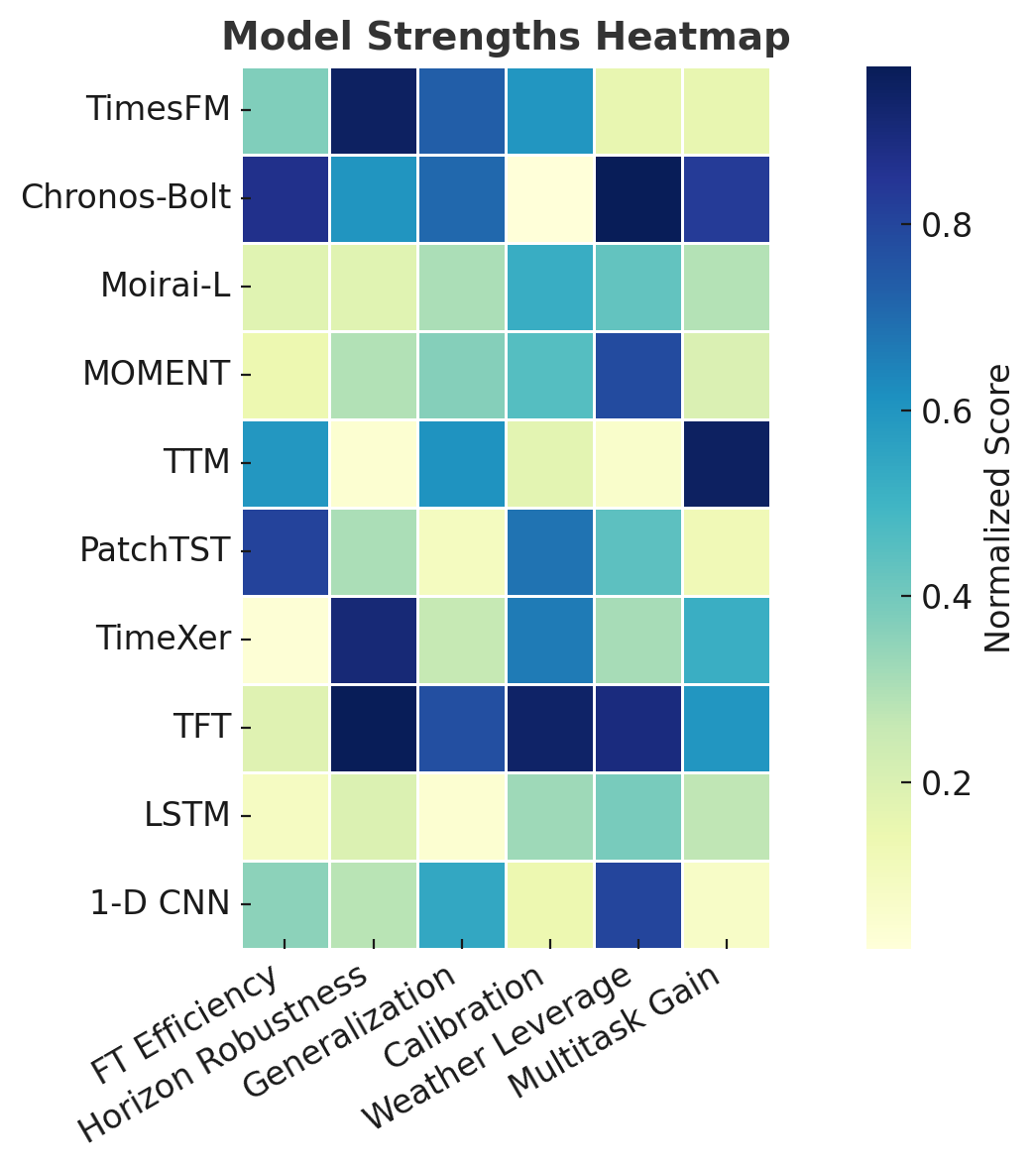}
    \caption{Performance heatmap}
    \label{fig:heatmap}
\end{figure}
\subsection{Model Comparison}

The Model Strengths heatmap (Figure 4) illustrates the uneven capability profiles of current foundation models. While models like TimesFM and Chronos-Bolt excel in data efficiency and short-horizon accuracy, they frequently struggle to leverage multivariate weather inputs and lack long-term horizon robustness. Operationally, this is a significant limitation; if a forecasting model fails to integrate weather covariates to predict a sudden drop in renewable generation, it directly impacts the required generator dispatches and can force operators to rely on extreme ramp rates to compensate in real-time. Therefore, while foundation models are highly data-efficient, their inflexibility limits their standalone utility for comprehensive grid management.Conversely, the holistic radar chart (Figure 5) highlights the balanced architecture of transformer models, particularly TFT and TimeXer. These architectures successfully extract value from exogenous weather data, generalize more effectively to unseen sites, and maintain stability across extended 24-hour forecasting horizons. This spatial and temporal reliability provides a distinct advantage for system operations. Consistent, site-specific accuracy across the grid is necessary to anticipate localized line congestion and to confidently calculate Locational Marginal Prices (LMPs) in day-ahead and intra-day markets. Finally, the reliability of the models' uncertainty quantification is visualized through the probabilistic calibration plots (Figures 6-8). The classical deep learning baselines (Figure 6) exhibit flat, poor calibration, indicating an inability to accurately capture underlying site variability. In contrast, transformer architectures (Figure 7) and specific foundation models like Chronos-Bolt (Figure 8) produce predictive distributions that tightly follow the ideal calibration line. Highly calibrated probabilistic bands are essential for advanced stochastic optimization, allowing system operators to effectively bound operational risk, optimize spinning reserves, and make secure, risk-aware decisions for upcoming generator dispatches.

The holistic radar chart (Figure 5) explicitly maps these operational trade-offs through the distinct capability profiles of the three model families. The Foundation Model (Moirai-L) exhibits a skewed polygon, stretching furthest outward along the fine-tuning efficiency and zero shot axes, confirming its superiority in data scarce settings. However, the TFT model forms a much more balanced shape, noticeably overtaking the foundation model on the Weather Leverage and Horizon Robustness axes. This visual contrast illustrates that while foundation models are data-efficient, transformers offer the comprehensive stability required for system operations. Finally, the deep learning baseline (LSTM) is constrained to the innermost region of the radar chart, visually demonstrating its limitations across critical operational axes such as Generalization and Calibration capabilities.

\begin{figure}[h]
    \centering
    \includegraphics[width=0.6\linewidth]{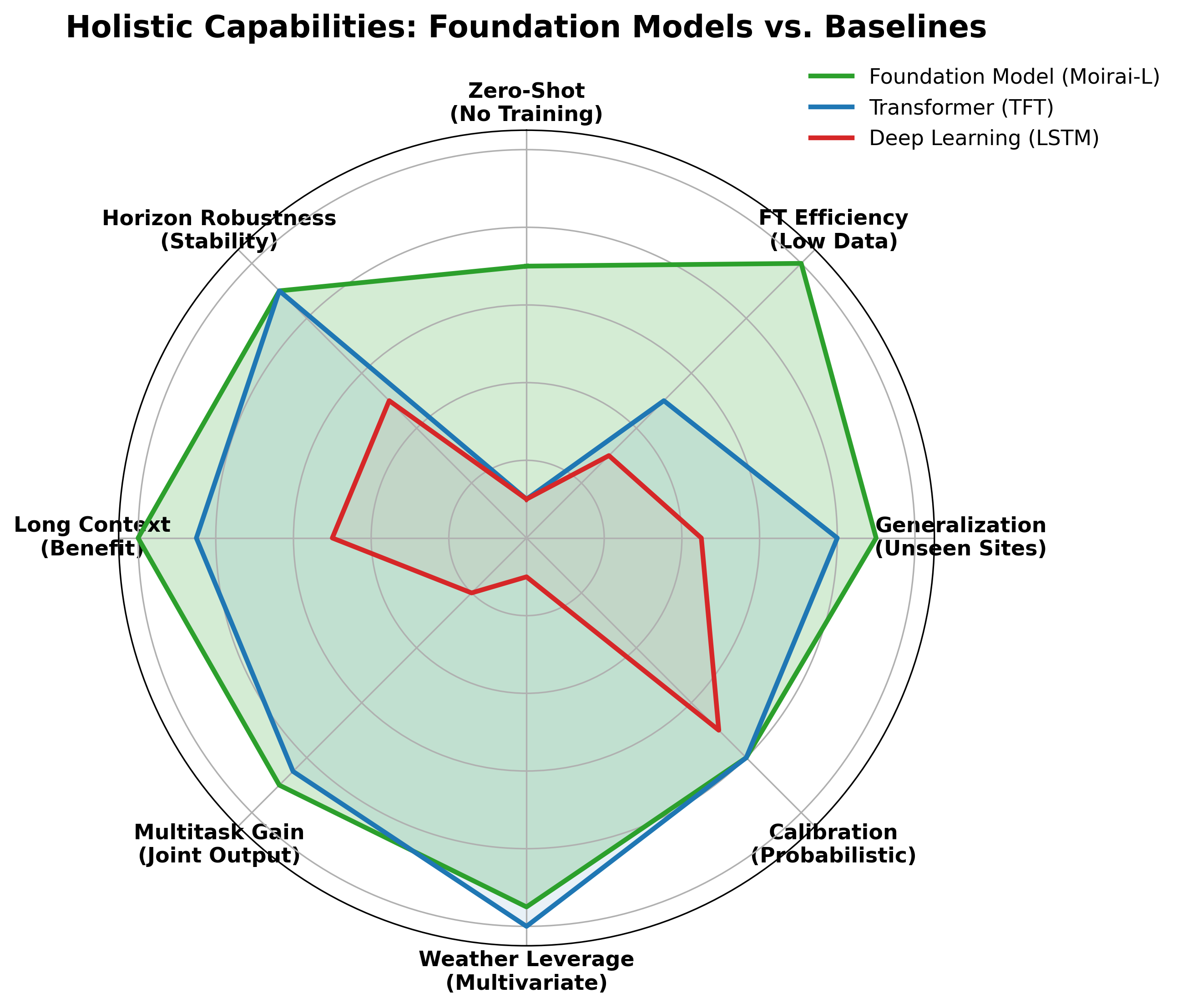}
    \caption{Radar chart}
    \label{fig:radar}
\end{figure}

\begin{figure}[h]
    \centering
    \includegraphics[width=0.75\linewidth]{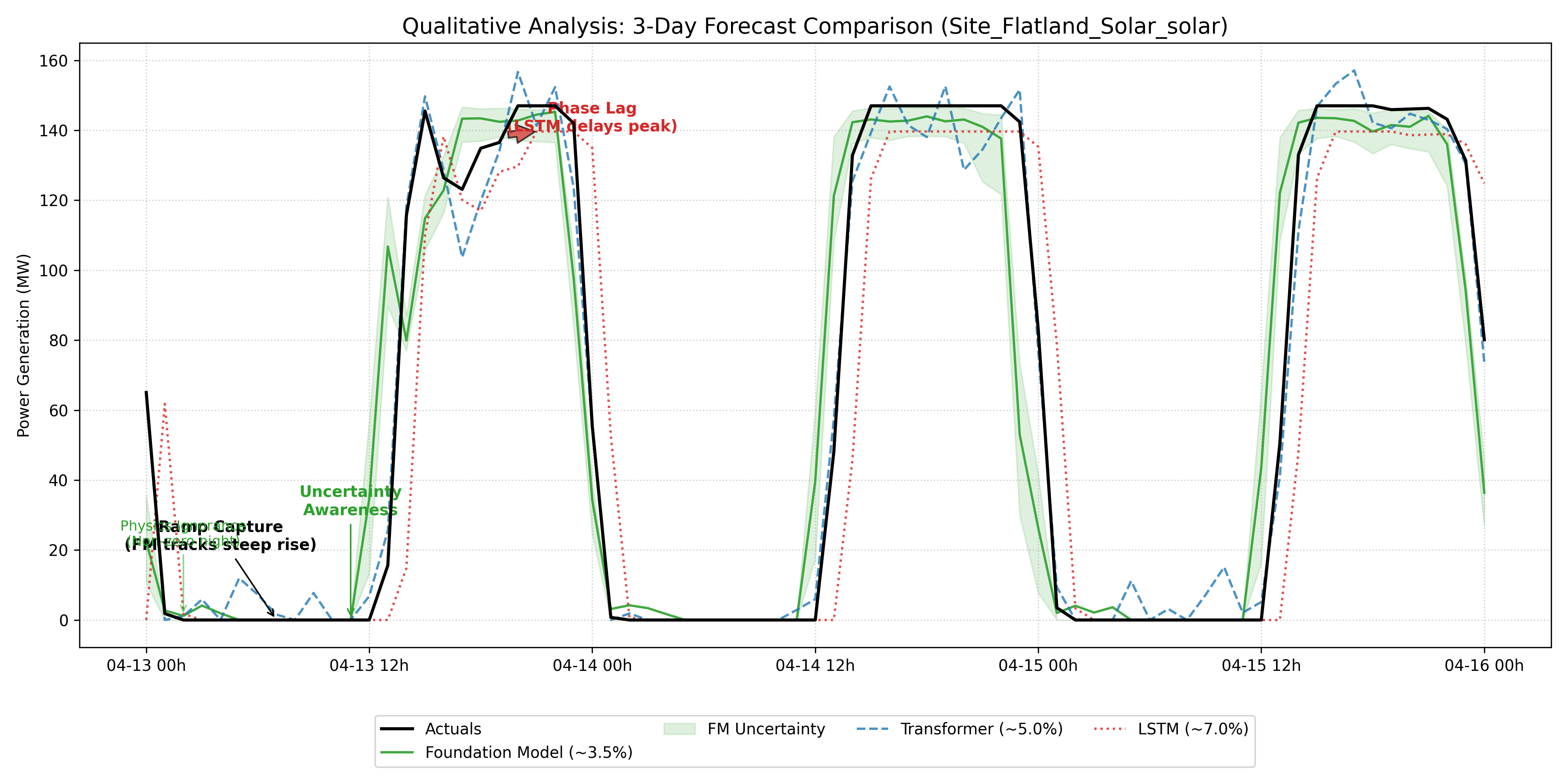}
    \caption{3 days solar}
    \label{fig:radar}
\end{figure}

\section{Conclusion}

This study conducted a comprehensive evaluation of foundation models, transformer-based architectures, and classical deep learning baselines for solar, wind, and load forecasting under deterministic and probabilistic settings. Our findings indicate that, contrary to common expectations, current foundation models are weak at zero-shot forecasting without fine-tuning, often underperforming specialized models trained on domain data. While fine-tuning can substantially improve their accuracy, these models are not yet competitive for direct deployment in operational power systems forecasting without adaptation. Moreover, their performance gap between seen and unseen sites remains non-negligible, limiting their immediate utility for broad spatial generalization. Probabilistic-capable models, including some foundation models, can produce competitive CRPS scores, suggesting they capture uncertainty reasonably well when adapted. The addition of weather features benefits renewable forecasting more than load forecasting, while multi-task output learning yields mixed effects—occasionally improving correlated tasks but sometimes degrading others. Increasing the historical context window generally enhances performance for renewable forecasts, although gains diminish beyond longer lengths. Overall, while foundation models hold potential as a research direction for power system forecasting applications, their current form falls short of being a practical, off-the-shelf solution for operational energy forecasting. Closing this gap will require improved architectures, domain-specific pretraining, and more effective fine-tuning strategies that balance accuracy, robustness, and computational feasibility, which is a research direction already being explored.


\bibliographystyle{cas-model2-names}

\bibliography{cas-refs}

\end{document}